\documentclass{aa}
\usepackage{epsf}

\def\beq#1{\begin{equation}\label{#1}}
\def\eeq{\end{equation}}
\def\beqa#1{\begin{eqnarray}\label{#1}}
\def\eeqa{\end{eqnarray}}

\def\comment#1{\relax}

\def\spose#1{\hbox to 0pt{#1\hss}}
\def\simlt{\mathrel{\spose{\lower 3pt\hbox{$\mathchar"218$}}
     \raise 2.0pt\hbox{$\mathchar"13C$}}}
\def\simgt{\mathrel{\spose{\lower 3pt\hbox{$\mathchar"218$}}
     \raise 2.0pt\hbox{$\mathchar"13E$}}}
\def\simpropto{\mathrel{\spose{\lower 3pt\hbox{$\mathchar"218$}}
     \raise 2.0pt\hbox{$\propto$}}}

\makeatletter
\def\eqalign#1{\null\,\vcenter{\openup\jot\m@th
  \ialign{\strut\hfil$\displaystyle{##}$&$\displaystyle{{}##}$\hfil
      \crcr#1\crcr}}\,}
\def\eqalignleft#1{\null\,\vcenter{\openup\jot\m@th
  \ialign{\strut$\displaystyle{##}$\hfil&$\displaystyle{{}##}$\hfil
      \crcr#1\crcr}}\,}
\makeatother

\begin{document}
\thesaurus{01      
      (08.14.1; 
       08.09.2: Her X-1)} 

\title{On the origin of X-ray dips in Her X-1}

\author{N.I.Shakura\inst{1}, M.E.~Prokhorov\inst{1},
K.A.~Postnov\inst{2}
\and N.A.Ketsaris\inst{1}}

\institute{Sternberg Astronomical Institute, Moscow University,
                119899 Moscow, Russia
\and
Faculty of Physics, Moscow State University,
                119899 Moscow, Russia
}

\date{Received 23 September 1998, accepted ..., 1998}
\maketitle
\markboth{N.I.Shakura et al. X-ray dips in Her X-1}{
...}

\begin{abstract}

A strong X-ray illumination of the optical
star atmosphere in Her X-1,
asymmetric because of a partial shadowing by the
tilted twisted  accretion disk around central neutron star,
leads to the formation of matter flows coming out of
the orbital plane and crossing the line of sight
before entering the disk. We suggest that
the absorption of X-ray emission by
this flow leads to the formation of pre-eclipse and
anomalous dips of type I. These dips are observed during
several orbits after turn-on both in the main-on and short-on state.
Almost coherent action of tidal torques and matter streams
enhances the disk wobbling which causes the disk edge to shield
the X-ray source after the turn-on. Anomalous dips of type II
and post-eclipse recovery appear due to this process only on the first
orbit after turn-on.

\end{abstract}

\section{Introduction}

Her X-1 is an accretion-powered 1.24-s X-ray pulsar in a binary system
with 1.7-d orbital period (Tananbaum et al. 1972).
The source displays 34.85-day X-ray intensity
variation due to eclipse by a counter-orbitally precessing
tilted accretion disk around the central neutron star
(Gerend \& Boynton 1976).
The X-ray light curve of Her X-1 consists of a main-on X-ray
state with a mean duration of $\sim 7$ orbital periods surrounded by two
off-states (also called low-on states) each of $\sim 4$ orbital cycles, and of a
smaller-intensity short-on state with a typical duration of
$\sim 5$ orbits, and is certainly formed by periodic obscurations of the
X-ray source by the disk.

As was understood already shortly
after the beginning of studies of Her X-1, the accretion disk
may be twisted. During the counter-orbital precession of such a disk
the outer parts of the disk open the central X-ray source while the inner
parts of the disk occult the X-ray source (Boynton 1978). Moreover, a hot
rarefied accretion disk corona may exist around its central
parts.  This makes the ingress to and egress from main-on and short-on states
asymmetric. The opening of the X-ray source with a rapid increase
of X-ray intensity is accompanied  by a notable
spectral changes which evidences for the presence of a strong absorption,
whereas the decrease in X-ray intensity occurs more slowly
and without appreciable spectral changes
(Giacconi et al. 1973).

One of the intriguing observational facts is that the X-ray source
always turns on near orbital phases $\phi_{orb}\simeq 0.2$ or $0.7$.
Such a behaviour has been explained by Levine \& Jernigan
(1982) by the accretion disk wobbling twice the orbital period due to
tidal torques. Indeed, it is at these orbital phases that the disk
angle inclination changes most rapidly.

\begin{figure*}
\epsfxsize=\textwidth
\epsfbox{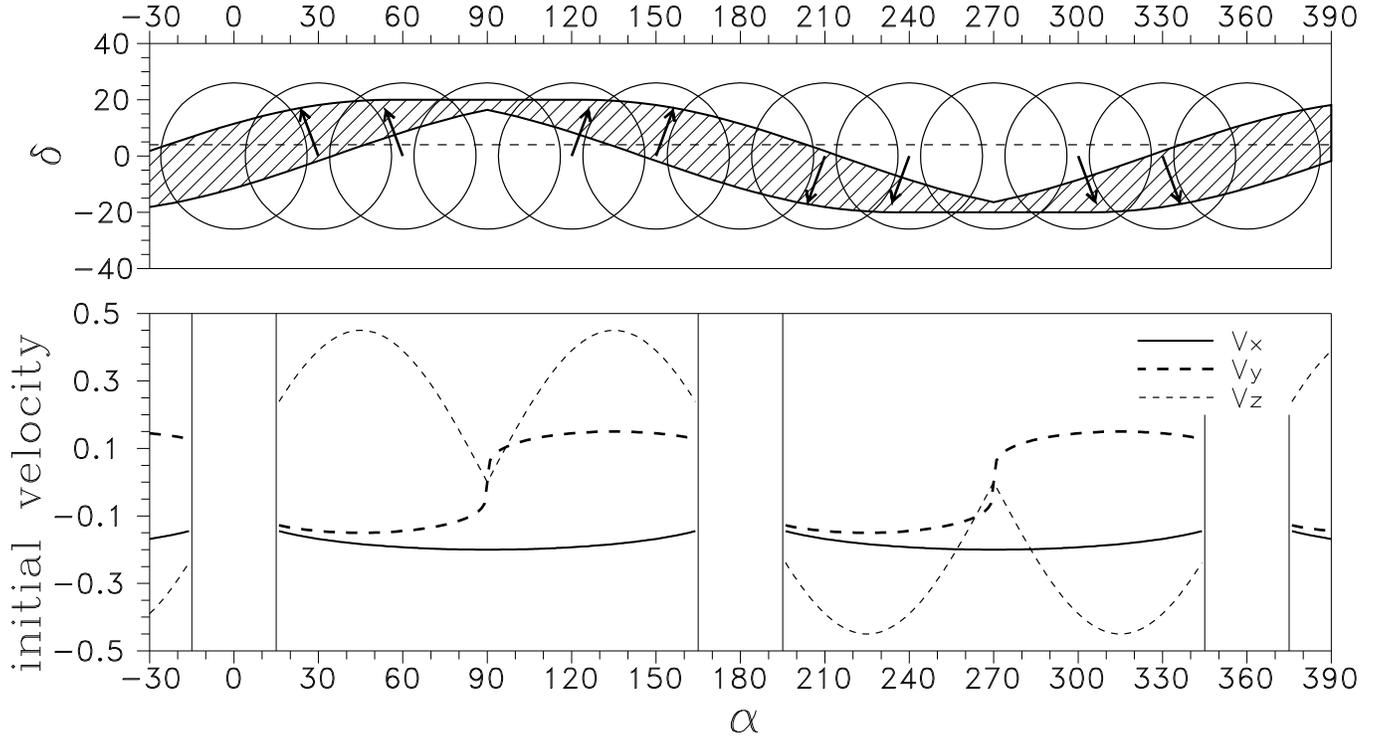}
\caption{Upper panel: the passage of HZ Her through the
shadow formed by a  tilted twisted
accretion disk. The coordinates $\alpha$ and $\delta$
count from the line of nodes of the middle part of the
disk along the orbital motion and from
the orbital plane, respectively. The dashed line indicates the
position of an observer inclined by $i=86^o$ to the
orbit. The arrows show the projection of the initial
accretion stream velocity on the plane
YZ perpendicular to the orbit. Bottom panel: The initial stream
velocity components at the $L_1$ point in units of the relative
orbital velocity 270 km/s. The flows disappear inside the shadow sector
between the outer and inner disk node lines.}
\end{figure*}

Another notable feature is that the duration of
successive 35-day cycles is as a rule 20, 20.5, or 21 orbital cycles
(Staubert et al. 1983). This behaviour has been confirmed by most
recent RXTE observations (Shakura et al. 1998).

Even more enigmatic features observed are sudden decreases in X-ray flux
(X-ray dips) which are accompanied by significant spectral changes. They
have been observed by many X-ray satellites: {\it UHURU} (Giacconi et al.
1973, Jones \& Forman 1976), {\it Copernicus} and {\it Ariel V} (Cooke \&
Page 1975, Davison \& Fabian 1977), {\it Ariel VI} (Ricketts et al. 1982),
{\it OSO-7, OSO-8} (Crosa and Boynton 1980), {\it HEAO-1} (Gorecki et al.
1982), {\it Tenma} (Ushimaru et al. 1989), {\it EXOSAT} (Reynolds \& Parmar
1995), {\it Ginga} (Choi et al. 1994; Leahy et al. 1994; Leahy 1997), {\it
RXTE} (Shakura et al. 1998; Scott \& Leahy 1998; Kuster et al. 1998). X-ray
dips are commonly separated into three groups: pre-eclipse dips (P), which
are observed in the first several orbits after X-ray turn-on (up to 7 in
main-on and up to 5 in short-on states) and march from the eclipse toward
earlier orbital phase in successive orbits; anomalous dips (A), which are
observed at $\phi_{orb}\sim 0.45-0.65$; post-eclipse recoveries (R), which
are occasionally observed as a short delay (up to a few hours) of the egress
from X-ray eclipse in the first orbit after turn-on.

Here we suggest a new model for all types of X-ray dips observed in Her X-1.
We shall show that the anomalous dips are separated into two types depending
on their formation mechanism. The key underlying feature of this model is
the anisotropic X-ray heating of the optical star atmosphere produced by the
tilted twisted accretion disk. This leads to the formation of gaseous
streams coming out of the orbital plane. Before entering the disk, the
streams may cross the observer's line of sight thus producing pre-eclipse
and anomalous dips (of type I). Such streams form the outer parts of the
accretion disk tilted with respect to the orbital plane. The tidal torques
causes the disk to precess in the direction opposite to the orbital motion
and also produce a notable wobbling (nutation) motion of the outer parts of
the disk twice the synodal period. Due to viscosity the disk becomes twisted
and all parts of the disk precess with one period, while the outer parts of
the disk undergo a notable wobbling caused by both tidal torques and
dynamical action of the streams. The X-ray source may be screened for some
time from the observer by the outer parts of such a disk in the first orbit
after the X-ray turn-on. Type II anomalous dips and post-eclipse recoveries
appear in this way.

\section{Origin of the pre-eclipse and type I anomalous dips}

In our model, a tilted twisted counter-orbitally precessing accretion disk
eclipses the central X-ray source between the main-on and short-on states.
The outer parts of this disk is inclined by an angle of $15-20^o$ relative
to the orbital plane. The inner parts of the disk are also tilted by
some angle, which can be determined from comparing the durations of
low-on states between the main-on and short-on states. As these low-on
states are, to an accuracy of 0.5 orbits, of the same duration (4 orbits),
the tilt of outer and inner parts of the disk should be comparable.

Such a disk produces an appreciable shadow and the
optical star periodically enters this shadow in its orbital motion (Fig.
1). The shadowed region is such that not all the optical star surface is
screened by the disk -- there always should exist areas illuminated by the
X-ray source with a photospheric temperature of 15,000-20,000 K whereas
photospheric temperature of the unheated regions is as low as $\sim 8,400$ K.
Even higher temperatures (up to $10^6$ K) due to soft X-ray absorption by heavy
elements are attainable in the chromospheric layers over the photosphere.
The X-ray heating of these layers is so strong that NV
$\lambda\lambda 1238.8, 1242.8$ doublet with a FWHM of 150 km/s is
observed (Boroson et al. 1996).

Such a high temperature induces matter outflow from the optical star which
would lie in the orbital plane in the absence of the shadow. The pioneering
calculations of such symmetric outflows were carried out in Basko \&
Sunyaev (1973). They showed that the induced mass outflow rate and
subsequent accretion on the neutron star is sufficient to explain the
observed X-ray luminosity $L_x\sim 10^{37}$ erg/s.

When the shadow appears, a powerful pressure gradients emerge in the
chromospheric layers near the boundary separating illuminated and obscured
parts of the optical star, which initiates large-scale motions of matter
near the inner Lagrangian point $L_1$ with a large velocity component
perpendicular to the orbital plane (Arons 1973; Katz 1973). Such a shadow
will periodically modulate the matter outflow rate $\dot M$, which is
dramatically reduced when the $L_1$ point is deep inside the shadow,
and rises rapidly to a maximum at the moment when the shadow edge
intersects the $L_1$ point. Clearly, the picture repeats twice over the
synodal orbital period.

Thus, matter flows non-coplanar with the orbital plane emerge and supply
the accretion disk with  angular momentum non-parallel to the orbital one.
Depending on the initial velocities, these streams may even increase the
disk tilt to the orbit. Such streams coming out of the $L_1$ point intersect
the line of sight of the observer at some orbital phases shortly before the
X-ray eclipse and shift slowly toward earlier phases as the precession
progresses. This is exactly the behaviour of the pre-eclipse X-ray dips
observed. The streams intersect the line of sight at other orbital phases
as well and thus give rise to the type I anomalous dips.

The problem of matter outflow from an asymmetrically illuminated stellar
atmosphere is essentially three-dimensional and requires sophisticated
numerical calculations. To obtain the pre-eclipse dips in a simplified
model, we calculated non-planar ballistic trajectories of particles ejected
from the point $L_1$. As a first step, we exploit some trial functions for
the initial outflow velocity components outside the shadowed sector (with
the optimal width being $\pm 15^o$ from the mean disk node line; see Fig.~2
and Table 1). Clearly, the sector lies within the region restricted by the
line of nodes of the outer parts of the disk on one side, and by the line of
nodes of the inner parts of the disk on the other side. The angle between
these lines is $70^o$ (see Table 1).

In a corotating frame with the X-axis directed from the X-ray source to the
optical star, the Y-axis pointing along the orbital motion, and the Z-axis
perpendicular to the orbital plane,
$v_{x}=-v_x^o|\sin\alpha|^{n_x}$,
$v_{y}=\mp v_y^o|\sin 2\alpha|^{n_y}$,
$v_{z}=\pm v_z^o|\sin 2\alpha|^{n_z}$,
with the angle $\alpha$ counted along the orbit from the mean accretion disk
node line (Fig.~2). The choice of the sign of $v_{y,z}$ is dictated by the
X-ray illumination conditions (see Fig.~1,2). The different angular
modulation of velocities comes from the following arguments. The shadow's
boundary crosses the $L_1$ point four times per orbital cycle, so the
powerful pressure gradients perpendicular to the shadow's boundary must
appear four times per orbital cycle. At the same time, the maximum outflow
rate along the X-coordinate is attained when the irradiation is maximal i.e.
when the shadow's boundary is far from the $L_1$ point, i.e. two times per
orbital period.

The initial velocities of particles were taken so that, on the one hand, the
dynamical action of the streams on the accretion disk keep it tilted by some
angle to the orbit, and on the other hand, the calculated dips occupy the
region of the observed dips on $\phi_{orb}-\phi_{pr}$ diagram. We set
$\phi_{orb}=0$ corresponding to the middle of X-ray eclipse and
$\phi_{pr}=0$ at the moment of maximal outer disk opening. On
$\phi_{orb}-\phi_{pr}$ diagram, however, it is more convenient to
plot $\phi_{pr}$ in terms of the number of orbital cycles
after X-ray turn-on at $\phi_{orb}=0.75$.

\begin{figure}
\epsfxsize=\columnwidth
\epsfbox{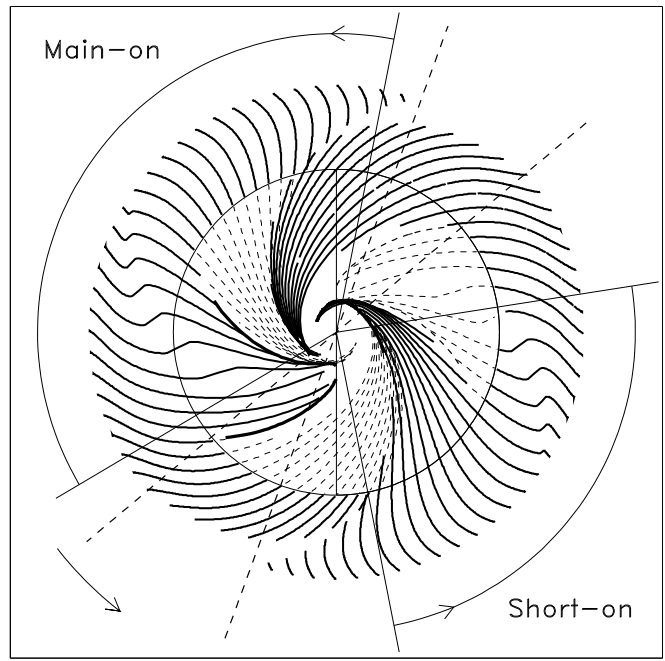}
\caption{The projection of the accretion streams onto the orbital plane
in the disk reference frame in $5^o$ orbital phase intervals.
The orbital motion is counter-clockwise.
The central circle indicates the projection of the outer parts of
twisted
accretion disk intersecting
the orbit along the vertical node line, the disk being above the
orbit to the right of this line. The inner Lagrangian point is shadowed by
the disk within the sector shown by the dashed lines which is restricted
by the node lines of the outer and inner parts of the disk.
Matter outflow is almost absent inside
this cone. The left and right wide sectors shown by solid lines correspond
to the main-on and short-on X-ray states,
respectively. The thick lines inside the accretion disk show the points
where the streams meet the disk plane. The dashed parts
of the streams mean they pass under the disk. The observer is
above the orbital plane. The bended structure of some stream lines
is due to angle modulation of $v_z^o$ velocity component.}
\end{figure}

The parameters of the model are summarized in Table 1. Note that with a
given disk tilt of $20^o$ and X-ray state durations the binary inclination
is unambiguously $i_b=86^o.0$. In principle, the calculations can be done
with the disk tilt $15^o$. Then the binary inclination would be $\approx
87^o$. This strict dependence is dictated by the relation between the
main-on, short-on, and low-on durations (see Table 1). From independent
analyses of optical light curves, the disk radius $R_d$ lies between $\simeq
(0.2-0.3)a$ (Gerend \& Boynton 1976, Howarth \& Wilson 1983, Shakura
et al. 1997). In the present paper we have taken $R_d=0.3 a$.

\begin{table*}
\caption{Model parameters}
\begin{tabular}{ll}
\hline
\hline

\multicolumn{2}{c}{Binary parameters}\\
Neutron star mass, $M_x$ & 1.4 $M_\odot$  \\
Optical star mass, $M_o$ & 2.2   $M_\odot$  \\
Orbital separation, $a$  & 9.1 $R_\odot$   \\
Orbital period, $P_{orb}$ & $1^d.7$\\
Relative orbital velocity, $v_{orb}$ & 270 km/s \\
Orbital inclination & $86^o.0$        \\

\multicolumn{2}{c}{X-ray light curve parameters}\\
Mean precession period, in $P_{orb}$ & 20.5\\
Main-On duration, in $P_{orb}$& 7.5\\
Short-On duration, in $P_{orb}$& 5.0\\
Low-On between Main-On and Short-On, in $P_{orb}$& 4\\
Low-On between Short-On and Main-On, in $P_{orb}$& 4\\

\multicolumn{2}{c}{Accretion disk parameters}\\
Disk radius, $R_d$ & $0.3 a$ \\
Outer disk radius tilt, $\theta_{o}$ & $20^o$ \\
$^\S$ Inner disk radius tilt, $\theta_{i}$ & $20^o$  \\
Angle between the line of nodes
of inner and outer disk, $\alpha_o-\alpha_i$ & $+70^o$\\

\multicolumn{2}{c}{Accretion stream parameters}\\
$^\dag$ Initial velocity components at the
inner Lagrangian point, in $v_{orb}$
& $v_{x}=-v_x^o|\sin\alpha|^{n_x},\qquad\;\,  v_x^o=0.20, \qquad n_x=0.25$ \\
& $v_{y}=\mp v_y^o|\sin 2\alpha|^{n_y},\qquad v_y^o=0.15, \qquad n_y=0.25$ \\
& $v_{z}=\pm v_z^o|\sin2\alpha|^{n_z},\qquad v_z^o=0.45, \qquad n_z=1$ \\
Width of the shadow sectors with no matter outflow, $W$ & $\pm 15^o$\\
Position of the shadow sectors bissectrix & $\alpha=0^o,180^o$\\
Minimal distances between the stream centre and the line of sight& 0.02$a$,
0.04$a$, 0.08$a$\\
\hline

\multicolumn{2}{c}{Resulting disc precession characteristics}\\
Mean tidal precession rate, $\Delta\alpha_t$ & $-42^o.2$ per sideral orbit\\
$^\ddag$ Mean dynamical precession rate, $\Delta\alpha_d$ & $+24^o.6$ per
sideral orbit\\

\hline
\hline
\multicolumn{2}{l}{\S
The difference in the inner and outer disk inclination
is determined by the difference
between the duration}\\
\multicolumn{2}{l}{of the low-on
states, which are equal to each other to an accuracy of 0.5 orbits.}\\

\multicolumn{2}{l}{\dag The angle $\alpha$ counts along the orbital motion
from the mean line of the nodes of the accretion disk.}\\

\multicolumn{2}{l}{\ddag  The total precession rate $\Delta\alpha=\Delta\alpha_t+\Delta
\alpha_d=-17^o.6\approx 360^o/20.5$}\\
\end{tabular}
\end{table*}

In a wide range of velocities $v_i^0$  and parameters $n_i$ the stream
crosses the line of sight before the X-ray eclipse and near orbital phases
$0.45-0.65$. We found a good coincidence for the initial velocity components
(in units of the orbital velocity $v_{orb}=2\pi a/P_{orb}\simeq 270$ km/s)
$v_x^0 \simeq 0.2$, $v_z^0 \simeq 0.45$, $v_y^0\simeq 0.15$.
The calculations
were performed for $n_x=n_y=0.25$, $n_z=1$. Note that the initial
$z$-component of the stream velocity is about two times higher than
its $x$-component.
This is due to large pressure gradients inside the
chromosphere near the inner Lagrangian point that appear
close to the boundary between the illuminated hot region ($T\sim 10^6$ K)
and the shadowed cool region ($T\sim 10^4$ K).
We should emphasize that the initial stream velocity
does not exceed the sound velocity in the hot region.
We also note that small perturbations of the velocity field
near $L_1$ point can result in sizable inhomogeneities of the
outflowing stream, so the pre-eclipse dip formed by
such a stream can have a complex structure.

Before colliding with the disk non-planar streams intersect the line
connecting the observer and central X-ray source thus absorbing some X-ray
flux. We identify these events with the pre-eclipse and type I anomalous
dips. In Fig.~3 we plot the calculated and observed dip positions on the
$\phi_{orb}-\phi_{pr}$ plane (data from Crosa \& Boynton (1980), Ricketts
et al. (1982), Shakura et al. (1998)). The contours include the calculated
dip positions for different minimal distances between the stream centre and
the line of sight (0.02, 0.04, and 0.08). The stream generated when the star
again enters the X-ray illuminated sector (see Fig.~2), must also intersect
the line of sight during the short-on state
$\phi_{pr}\approx 0.25-0.55$. This gives rise to the pre-eclipse dips
during the short-on state as well, as indeed observed (Jones \& Forman
1976; Ricketts et al. 1982; Shakura et al. 1998).

Fig.~3 demonstrates qualitative agreement between the observed and
calculated dip positions. Note that in our model parameters additional
pre-eclipse X-ray dips can appear at the end of the "main-on" state
when the stream from another illuminated sector crosses the line of
sight (see Fig.~2,3). It would be interesting to look for these
features in detailed X-ray observations. Interestingly, with the
parameters assumed, in the first several orbits the pre-eclipse dip is
slightly separated from the X-ray eclipse.

Note that a small difference in the calculated and observed anomalous dip
positions in the main-on state (Fig.~3) is due to the roughness of the
model. The agreement can be made better by shifting the maxima of $v_y$,
$v_z$ velocities by $\sim 10-15^o$ towards the shadow boundaries. For
example, one may take gaussian-like trial functions for phase velocity
dependence. However, this is hardly worthwhile doing until more detailed
X-ray dip observations become available.

\begin{figure}
\epsfxsize=\columnwidth
\epsfbox{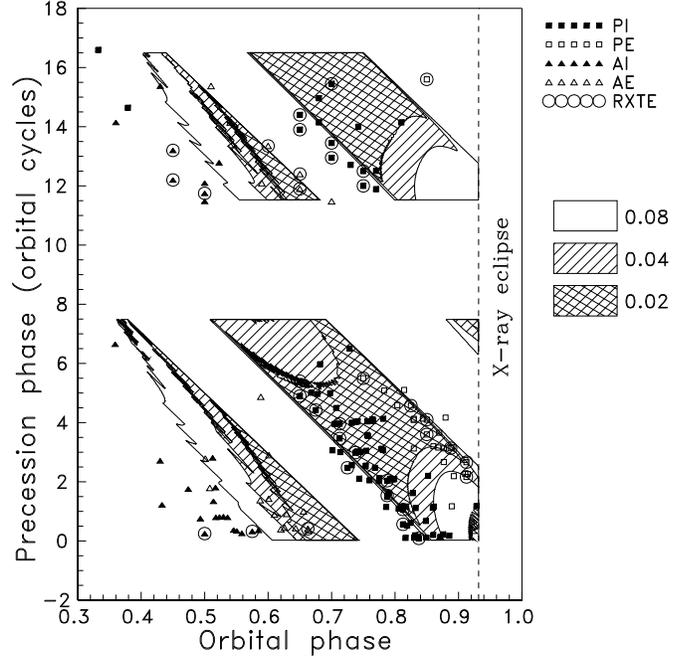}
\caption{The part of
the plane $\phi_{orb}-\phi_{prec}$ with the observed
pre-eclipse and anomalous
X-ray dips in the main-on and short-on states. Quadrangles and
triangles indicate ingress to (PI) and egress from (PE) the pre-eclipse dips.
RXTE data are encircled.
The calculated pre-eclipse dips and short anomalous dips arise when
the accretion stream intersects the line between the X-ray source and the
observer before entering the disk. Different contours correspond to
different minimal distances (0.02, 0.04, 0.08) between the stream centre and
the line of sight.}
\end{figure}

\section{Features of the precessional motion in Her X-1}

The motion of the tilted accretion disk proceeds, on the one hand,
under the tidal action from the optical star, and, on the other hand,
suffers from the dynamical action of the gaseous streams. The total
action of these torques makes it precess with a period of 20.5 orbital
cycles (the 35-day X-ray cycle).

First we calculate the tidal torque applied to a ring of matter
of radius $r$ at the outer edge of the disk.
\begin{equation}
\frac{d\vec K_t}{dt}=\int_0^{2\pi} [\vec r\times \vec f_t(\phi)]\,d\phi\,,
\label{dKdt}
\end{equation}
where $\vec f_t(\phi)$ is the tidal force applied to the ring's
element.
Under the action of this torque
the mean counter-orbital precessional motion proceeds and the
wobbling at twice the orbital period takes place both in
angles $\alpha$ and $\theta$.

\subsection{Mean precession rate}

We calculated the rate of the mean
precessional motion for such a ring and compared it with the
well-known analytical expression obtained in quadrapole approximation
\begin{equation}
P_{pr}^t\simeq \Pi_0\Pi_1
\frac{4}{3}\frac{(q+q^2)^{1/2}}{R_d^{3/2}\cos\theta}P_{orb}\,,
\label{Ppr}
\end{equation}
where $q\equiv M_x/M_o\simeq 0.64$ is the mass ratio, $P_{orb}=1.7^d$ is the
orbital period. We introduced here a numerical factor $\Pi_0< 1$ which can
account for the additional effects of higher multiples (see Table 2).

\begin{table}[h]
\caption{}
\begin{tabular}{lcl}
$R_d/a$ &\qquad& $\Pi_0$\\
0.3     &\qquad& 0.87 \\
0.25    &\qquad& 0.91 \\
0.2     &\qquad& 0.94 \\
0.15    &\qquad& 0.97 \\
0.1     &\qquad& 0.985 \\
\end{tabular}
\end{table}

In fact, we deal not with a rigid ring but with a viscous disk in which the
inner parts should precess much more slowly under the action of tidal
torques. However, due to viscosity the disk precesses as a whole with one
period which is slightly larger than the period of the outer parts. To
account for this effect, we introduced another numerical factor $\Pi_1\sim
1.1$. With the assumed parameters of Her X-1 ($q\simeq 0.64$, $R_d\simeq 0.3
a$, $\theta\simeq 20^o$), the pure tidal precession rate per sideral
orbital period would be $\sim -42^o.2$, whereas the observed rate is
$-17.^o6$. The difference between them is compensated by
the dynamical action of the streams.

One can calculate dynamical action of the streams on the disk by
solving the equation for the disk angular momentum $\vec K_d$ change:
\begin{equation}
\frac{d\vec K_d}{dt}=[\vec r\times \vec f]=\dot M_{in}[\vec r\times \vec v]\,.
\end{equation}
where $[\vec r\times \vec f]$ is the torque applied to the disk by a
stream with mass flow rate $\dot M_{in}$ colliding with the disk at
velocity $\vec v$ at a distance $\vec r$ from the disk centre. The
outflow rate is $\dot M_{out} \propto |\vec v|_{L_1}$. Clearly, a phase
lag $\Delta \alpha$ exists between the time of outflow of a given
portion of matter from the $L_1$ point and its encounter with the disk,
so $\dot M_{in}(\alpha)= \dot M_{out}(\alpha+\Delta \alpha)$, which we
take into account.
Although we cannot take into account the dependence of
the accretion rate on the matter density at the Lagrangian point,
nevertheless we are able to
introduce a dimensionless numerical factor such that the dynamical action of
the streams together with the tidal torques lead to the observed precession
rate $-17^o.6$ per sideral orbit. Thus the dynamical action makes the disk to
precess along with the orbital motion with the rate
$-17^o.6-(-42^o.2)=+25^o.6$ per sideral orbit. So the absolute value of the
contribution of the dynamical torques to the observed precession rate
is of order 50\% of tidal torques. For a disk with size $0.25 a$ the
tidal precession rate would be $-30^o.7$  and correspondingly to provide
the observed rate $-17^o.6$ per orbit the dynamical precession rate
should be $13^o.2$ per orbit, i.e the contribution of dynamical torques
would be $\sim 40\%$ of tidal torques.
If the disk radius is well defined
then the $\alpha$-parameter of disk accretion
can be evaluated by comparison of dynamical and tidal actions.

The streams supplying the disk with angular momentum  would change not only
its precession rate but also its tilt to the orbit. We found that for fixed
$v_x^0$, $v_y^0$ and $v_z^0$ the streams tend to increase a small disk tilt
and {\it vice versa}, if the tilt is not small, the streams would decrease
it, so that an equilibrium disk inclination angle $\theta$ can always be
found for given $v_i^0$. The viscous tidal effects tend to decrease the
equilibrium tilt by a factor
$\tau_t/(\tau_t+\tau_d)$, where $\tau_d\equiv M_d/\dot M$ and $\tau_t$ are
characteristic times for dynamical and viscous tidal actions, respectively.

Note that  with accretion rate $\dot M$ growth the dynamical torque
increases, while the tidal torque does not change (assuming the
constant disk radius). So the net precessional period should increase.
Indeed, such a correlation has been deduced between the duration of the
35-day cycle and the mean X-ray flux from analysis of observations by
BATSE (Wilson et al. 1994) and RXTE (Shakura et al. 1998).

\subsection{Wobbling of the outer disk parts}

In addition to the mean precession motion of the disk, there is a wobbling
of the outer parts of the disk due to both the tidal torques and dynamical
action of the streams. The pure tidal wobbling in quadruple approximation
was studied by Levine \& Jernigan (1982) who showed that this effect causes
the source to turn-on near the orbital phases 0.2 and 0.7.

When calculating the tidal wobbling we used the general Eq.(\ref{dKdt}).
Then we approximated the wobbling motion of the outer disk
by the following harmonic functions: for the angle of the line of
nodes of outer disk
$$
\Delta\alpha_t\equiv\alpha_o-\langle \alpha_o\rangle=A_2 \sin 2\alpha_o\,,
$$
for the tilt of the outer disk
$$
\Delta \theta_t\equiv \theta_o-\langle \theta_o\rangle
=B_2 \cos 2\alpha_o\,.
$$
Higher-order multiples contribute less than about 1\%,
i.e. the tidal wobbling follows almost exactly a pure sine law at the
double orbital frequency, while the quadruple approximation
for the mean precession  motion turns out to be more crude (see Table 2).

We also calculated numerically the wobbling due to dynamical action of the
stream with taking into
account the renormalization of the dynamical torques to the
observed precession rate per orbit discussed above. Since the stream torque
depends on the orbital phase in a complex way (unlike the tidal torque,
it is far from being sine-like), the resulting periodical variations of
$\Delta\alpha_o$ and $\Delta\theta_o$ with the orbital phase have very complex
shapes (see Fig.~4). In Fig.~4(a), we have plotted the tidal,
dynamical, and the total change in the outer disk tilt $\Delta\theta_o$
as a function of the phase angle of the mean line of nodes of the outer
disk $\alpha_o$.
Fig.~4(b) shows the same for the wobbling of the line of nodes of
the outer disk  $\Delta\alpha_o$.

To conclude this section, we note that in a general case
the  accretion rate $\dot M$ (and hence dynamical torques
applied to the disk)  may be phase-dependent in addition
to the dependence through the stream velocity $\vec v(\alpha)$:
 $\dot M_{out}\propto |\vec v|f(2\alpha)$, where
$f$ is some periodical function. This additional dependence has been
neglected in our calculations ($f\equiv 1$).

\begin{figure}
\epsfxsize=\columnwidth
\epsfbox{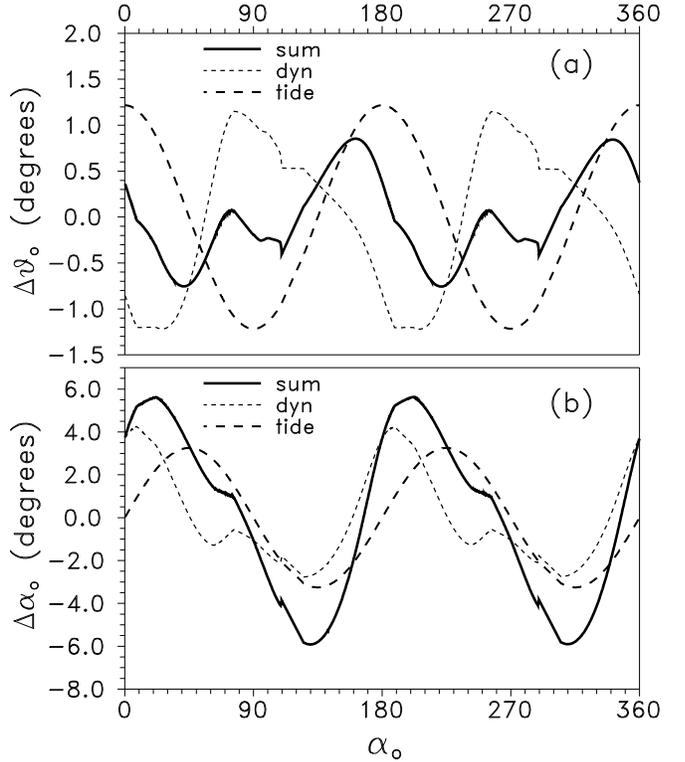}
\caption{
The wobbling variations of the outer disk tilt $\Delta\theta_o$
(a) and its line of the nodes $\Delta\alpha_o$ (b)
as a function of the phase angle $\alpha_o$.
}
\end{figure}

\section{Origin of post-eclipse
recoveries and type II anomalous X-ray dips}

Unlike the pre-eclipse dips and type I anomalous dips, the anomalous dips of
type II and post-eclipse recoveries are formed by another mechanism. The
vector of the disk angular momentum moves along a precession cone and
undergoes an oscillating (wobbling) motion twice the synodical orbital
period.

We recall that the secular precession effects due to dynamical and tidal
interactions have the opposite signs. As shown above, the oscillating part of
both torques behaves in a more complex way. So does the angle $\epsilon$
between the line of sight and the outer disk plane:
\begin{equation}
\cos\epsilon=\cos i \cos\theta_o + \sin i \sin \theta_o \cos \phi_{pr}\,.
\end{equation}
At the precession phases
where the outer disk is observed face-on, the variation in
$\epsilon$ is mostly due to changes in the angle $\theta_o$, which
are small. As seen from Fig.4(a), the deviations of angle $\theta_o$ caused
by tidal and dynamical actions are in counter-phase and practically
compensate for each other. In contrast, when the disk is observed
edge-on, the main contribution into change of $\epsilon$ comes from
the variations of $\alpha_o$, for which the dynamical and tidal
torques act almost coherently (Fig.~4(b)) enhancing the resulting change in
$\epsilon$.

Fig.~5 shows the angle $\epsilon$ between the line of sight and the
outer disk plane (the jagged curve). As seen from this figure, X-ray
turn-ons may be followed by a short time interval in which the X-rays
are again occulted by the outer edge of the disk. We identify this
effect with either a type II anomalous dip (A) at $\phi_{orb}\sim
0.4-0.6$) or a post-eclipse recovery (R) at $\phi_{orb}\simeq 0.1-0.2$.
 We also note that these features do not appear more than in one
orbital cycle after turn-on and moreover, unless the disk radius is
unrealistically large ($> 0.3 a$), only a post-eclipse recovery or an
anomalous dip of type II is observed.

\begin{figure}
\epsfxsize=\columnwidth
\epsfbox{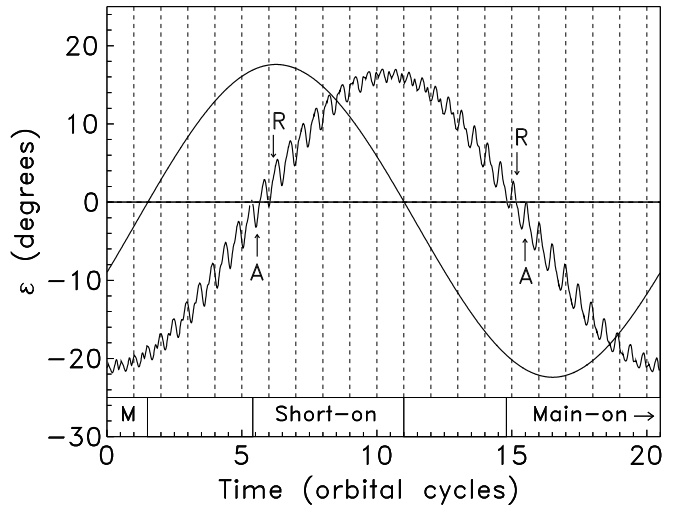}
\caption{
The angle $\epsilon$ between the line of sight and the outer accretion
disk plane (the jagged line) and inner accretion disk plane (the smooth
line) as a function of time (in orbital cycles).
The small vertical arrows indicate the anomalous X-ray dips of type II (A)
and post-eclipse recovery (R).}
\end{figure}

\section{Discussion}

We have suggested a model explaining absorption features (X-ray dips and
post-eclipse recoveries) in the X-ray curve of Her X-1. In this model,
anisotropic X-ray heating of the optical star atmosphere causes the matter
outflow from the inner Lagrangian point to proceed in the form of
non-coplanar streams. These streams intersect the line of sight of the
observer forming different absorption features (X-ray dips). The model
demonstrates a satisfactory agreement with observed positions of these
features on $\phi_{orb}-\phi_{pr}$ plane. The agreement can be made better
by choosing trial functions for the initial stream velocity components at
the inner Lagrangian points in a Gaussian-like form.

The X-ray observations of Her X-1 from {\it EXOSAT} (Reynolds and Parmar
1995) and especially from {\it Ginga} (Choi et al. 1994; Leahy et al. 1994;
Leahy 1997) satellites revealed that the absorption dips have different
duration and complex structure. In our model, we expect the pre-eclipse
X-ray dips and type I anomalous dips which are formed due to absorption in
the streams to have a complicated structure, while the type II anomalous dips
and post-eclipse recoveries, which are due to the outer disk wobbling, must
be more smooth. So the observation of energy distribution inside dips would
be very important and dedicated continuous X-ray observations covering the
whole 35-day cycle are highly desirable to clarify this picture.

The essential feature of our model is the presence of a twisted tilted
accretion disk around the central neutron star. The presence of such a disk
in Her X-1 binary has long been discussed in literature (Petterson 1977,
Schandl \& Meyer 1994,  see
detailed reference in Wijers \& Pringle 1998). We used a twisted disk in
which the outer parts are tilted to the orbit by the same amount as the
inner parts. The calculation made by Wijers \& Pringle (1998) have
suggested that the inner parts of the disk may even have a higher tilt than
the outer parts by the action of radiation pressure. When calculating the
twisted disk parameters ($\theta_o$, $\theta_i$, $\alpha_0-\alpha_i$) we
have come from the observed duration of the main-on (7.5 orbits), short-on
(5 orbits), and low-on (4+4 orbits) states in Her X-1. As the error in the
determination of these states may be noticeable (about 0.5 orbits), it is
not excluded that the inner parts of the disk may be much more tilted to the
orbit. For instance, if the durations of these states were related as 7
(main-on) : 4.5 (1st low-on) : 5.5 (short-on) : 3.5 (2d low-on), the inner
disk tilt would be $\sim 60^o$ with the twist angle
$\alpha_o-\alpha_i\sim 90^o$.
In our model of twisted accretion disk the twist angle could be
so high that the "windows" where main-on and short-on states
are observed may be "closed" (see Fig.~5). This can explain
the prolonged off-state observed during 9 months in 1983-1984
(Parmar et al. 1985).

A geometrical thick accretion disk with $H/R\sim 0.2$
would generally produce a similar
shadow on the optical star atmosphere as the twisted geometrically thin
disk. Since in our model it is the
shape of the shadow
that determines the initial velocity components, the behaviour of
X-ray dips can be explained by such a thick disk as well. But we do use
the thin twisted disk models because a very high temperature
is required to produce such a thick disk. Clearly, the choice of
the disk model requires additional studies and cannot be uniquely made
solely on the grounds of X-ray dip behaviour.
In principle, a more thorough study of the
optical light curves of HZ Her could put new bounds on the disk parameters.

\begin{acknowledgements}
We thank the referee Dr. Susanne Schandl for constructive notes.
The work was supported by the Grant "Universities of
Russia", No5559, Russian Fund for Basic Research through Grant No
98-02-16801 and the INTAS Grant No 93-3364-ext. N.I.S. acknowledge
the staff of Cosmic Radiation Laboratory of RIKEN (Tokyo) and
Max-Planck Institut f\"ur Astrophysik (Garching) for hospitality.
\end{acknowledgements}

\end{document}